# Deep Geometry Handling and Fragment-wise Molecular 3D Graph Generation

Odin Zhang[1, #], Yufei Huang[2, #], Shichen Cheng[1, #], Mengyao Yu[1, #], Xujun Zhang[1], Haitao Lin[2], Yundian Zeng[1], Mingyang Wang[1], Zhenxing Wu[1], Huifeng Zhao[1], Zaixi Zhang[3], Chenqing Hua[4], Yu Kang[1], Sunliang Cui[1, *], Peichen Pan[1, *], Chang-Yu Hsieh[1, *], Tingjun Hou[1, *]

[1]College of Pharmaceutical Sciences, Zhejiang University, Hangzhou 310058, Zhejiang, China
[2]Zhejiang University, Hangzhou 310058, Zhejiang, China
[3]Anhui Province Key Lab of Big Data Analysis and Application, University of Science and Technology of China, Hefei, Anhui, China
[4]Montreal Institute for Learning Algorithms, McGill University, Montreal, QC, Canada

[#]Equivalent authors

**Corresponding authors**
    **Tingjun Hou**
    **E-mail:** tingjunhou@zju.edu.cn
    **Chang-Yu Hsieh**
    **E-mail:** kimhsieh@zju.edu.cn
    **Peichen Pan**
    **E-mail:** panpeichen@zju.edu.cn
    **Sunliang Cui**
    **E-mail:** slcui@zju.edu.cn

## Abstract

3D structure-based molecular graph generation has attracted tremendous attention for its vast potential in designing potent drug candidates. Most earlier approaches follow an atom-wise paradigm, incrementally adding atoms to a partially built molecular fragment within protein pockets. These methods, while effective in designing tightly bound ligands, often overlook other essential properties such as synthesizability and drugability. The fragment-wise generation paradigm offers a promising solution by assembling chemically sensible fragments to reduce synthesis difficulty. However, a common challenge across both atom-wise and fragment-wise methods lies in their limited ability to co-design plausible chemical and geometrical structures, resulting in distorted conformations and inferior binding tendencies. In response to this challenge, we introduce the Deep Geometry Handling protocol, a more abstract design that decomposes the entire geometry into multiple sets of geometric variables, extending the design focus beyond the model architecture. Through a comprehensive review of existing geometry-related models, we identify and discuss six protocols, each with its own strengths and weaknesses. Building upon these insights, we propose a novel hybrid strategy, culminating in the development of FragGen – the first geometry-reliable, fragment-wise molecular generation method. FragGen marks a significant leap forward in the quality of generated geometry and the synthesis accessibility of molecules—addressing two major challenges in the application of molecular generation algorithms. The efficacy of FragGen is further validated by its successful application in designing type II kinase inhibitors at the nanomolar level, grounding its algorithmic advancements in real-world drug development. We believe that this concept-algorithm-application loop will not only inspire researchers working on other geometry-centric tasks to look beyond architecture design, but also offer a concrete example of applying structure-based generative artificial intelligence in drug design.

# Introduction

Despite the emergence of a plethora of novel modalities in the past decade, designing druggable molecules that target functional protein remains the most effective treatment option. Empowered by the rapid advancement of artificial intelligence (AI)-aided drug design (AIDD)[1], our ability to discover suitable organic-molecule-based drug candidate has been dramatically enhanced. The ambitious endeavor of computer-aided drug discovery primarily bifurcates into two streams: virtual screening, which involves sifting through existing molecular libraries[2], and molecular generation, which entails crafting molecules from scratch[3]. The former, essentially a classification task, has seen significant development over the past decade in the AI landscape, exemplified by advancements in scoring functions[4]. On the other hand, the later has been synergized with the language and graph generation methods, leading to SMILES-based[5] and graph-based molecular generation models[6], bringing in fresh computational perspectives to drug discovery. Despite the progress in AIDD, the absence of any AI-designed drugs passing regulatory approval highlights the formidable challenge of data-driven drug design. A key issue is data sparsity, a domain-specific obstacle that does not severely affect other fields like image or language processing where extensive data is available. In drug discovery, limited datasets are common due to the high costs and complexity of drug development, confidentiality in pharmaceutical research, and the vastly complex functioning principles of biological systems[7]. Data scarcity restricts the potential and applicability of many advanced AI models that have previously been proven successful in data-rich environments. Thus, external assistance, particularly in the form of physical constraints, becomes crucial to mitigate this intrinsic challenge by introducing prior knowledge to restrain the solution space. The rapid development and impressive performance of AlphaFold[8] and other structure-related models[9] underscore the efficacy of this approach. Concurrently, there is a growing emphasis on structure-based methodologies in both virtual screening and molecular generation, opening up new frontiers and challenges, such as binding conformation prediction[10] and pocket-aware molecular generation[11].

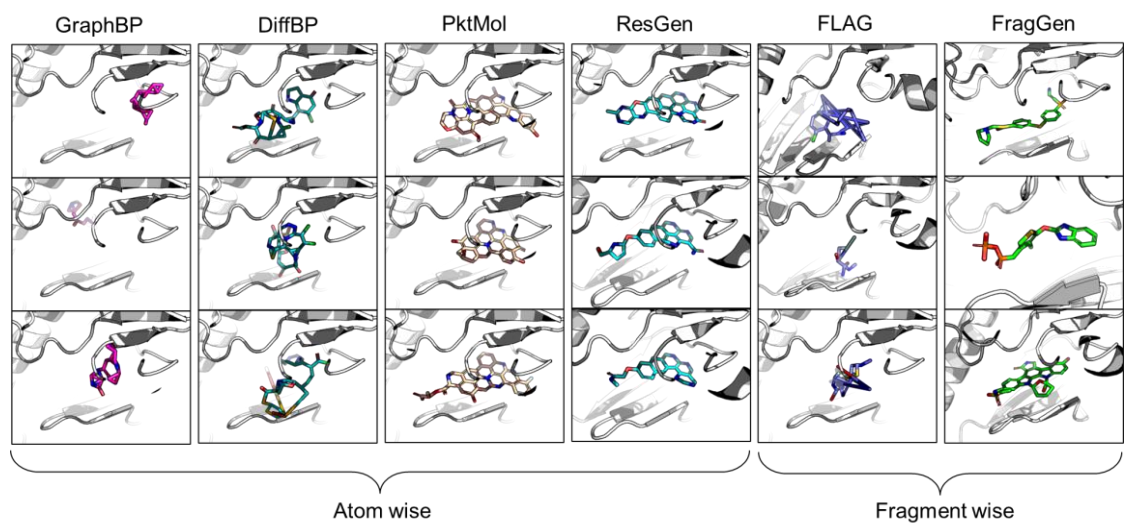

**Figure 1.** Visualized molecules generated by different methods. All models are performed without force-field optimization.

In the realm of 3D pocket-aware molecular generation, recent years have witnessed the emergence of many promising models like LiGAN[12], Pkt2Mol[13], DiffBP[14], ResGen[15], etc, which have manifested varying degrees of success in generating potentially superior ligands with a lower binding energy (as estimated by docking scores) than the reference ligands. However, a closer inspection on the generated ligands, particularly before any post-processing, reveals two critical limitations of most existing models. Firstly, the generated molecular conformations often appear distorted, which is noted in the outputs of GraphBP[16] and DiffBP (**Figure 1**). Secondly, there is a tendency to produce molecules with multi-fused rings to fill the cavity of protein pockets, which is observed in the outputs of Pkt2Mol and ResGen (**Figure 1**). While these generated structures may induce stronger interactions with protein pockets, they either look physically implausible or the complex structure poses significant challenges in synthesis and often results in toxic properties, thus actually distancing them from ideal drug candidates. Fragment-wise molecular generation offers a solution to the multi-fused ring issue by assembling a molecule in terms of synthesizable fragments as basic elements during the generative process. However, the only existing implementation of this approach, i.e., FLAG[17], encounters significant challenges with geometry handling as illustrated in **Figure 1**. The error in each fragment generation step accumulates, ultimately causing the collapse of the molecular structure. Therefore, there is a pressing need for a reliable fragment-wise deep generative model in structure-based drug design (SBDD).

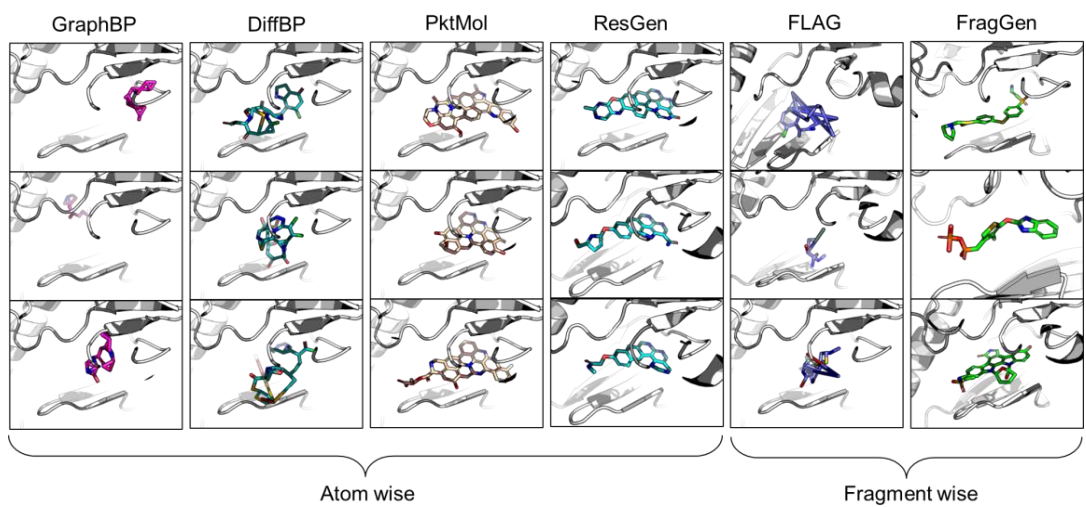

**Figure 1.** Visualized molecules generated by different methods.

Rendering smooth geometries is a central focus of computational study of physical reality, not just for 3D molecular generation but across almost all geometry-centric application domains. For instance, in molecular conformation generation, researchers [18] have adopted the distance-then-geometry protocol to first generate distance matrices then deduce Cartesian coordinates by optimizing randomly initialized conformations under the distance constraint. However, the non-uniqueness in mapping under-specified distance matrices to Cartesian coordinates often introduces additional errors, leading to geometric distortions. Subsequent research[19,20] has explored force-field optimization or end-to-end Cartesian coordinate prediction to enhance a deep learning model's capability to generate accurate geometry. In addition to efforts on the direct generations of plausible molecular conformations, deep learning has also concurrently made significant advancement on the front of molecular docking. Early models, such as TANKBind[21], extended the idea of distance-then-geometry protocol to protein-ligand binding conformation prediction. However, the incorporation of protein nodes into these models introduced a formidable challenge: a significant increase in redundant degrees of freedom, which led to unsatisfactory geometries. Then researchers delved into the end-to-end solutions, directly predicting the Cartesian coordinates, as pioneered by EquiBind[22]. KarmaDock[23] further advanced this protocol by employing a recycling mechanism, emulating the classical geometry optimization, and finally raising the successful rate of docking by about 50%. Yet, all these methods still struggle with the generation of unrealistic local structures, such as non-coplanar aromatic rings and excessively long chemical bonds, necessitating post-processing steps like geometry optimization or alignment corrections. DiffDock[24] represents a different technical approach, focusing on tuning constrained variables like overall translation, orientation, and torsion angles in order to simplify the morphing of molecular conformations. DiffDock's idea works well as it improves the state of geometric plausibility of deep-learning based generations, though its generated ligands may still encounter clashes with protein pocket residues.

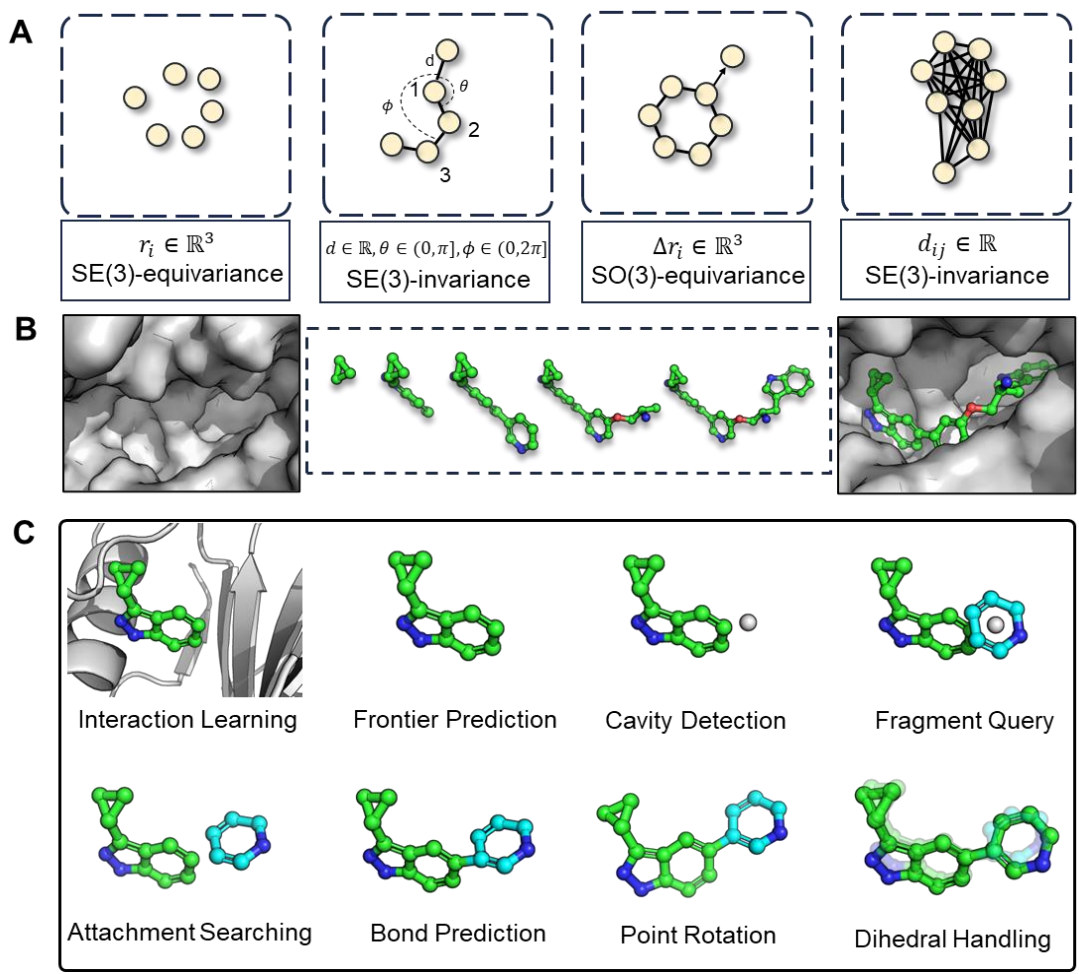

**Figure 2. A).** Illustration of symmetry requirements for various geometric variables. **B).** Structure-aware and fragment-wise molecular generation **C).** Workflow of our proposed combined geometry handling protocol, which is specifically designed for 3D fragment-wise molecular generation.

The challenges in correctly handling geometry with deep learning models are twofold: the inherent symmetries in geometric variables (illustrated in **Figure 2A**) and in which way the geometry is constructed. The first aspect, symmetry considerations, like SE(3)-invariance/equivariance, has been thoroughly addressed. Many works have concentrated on enhancing the feature extraction capability of models while enforcing adherence to the necessary equivariance or invariance principles. For example, the transformation of Cartesian coordinates should comply with roto-translational equivariance, which is mathematically expressed as Rf(x) = f(Rx), where R represents the rotation matrix and f denotes the neural network function. However, the second aspect, the high-level geometric handling protocol, has not received as much attention compared to the development of symmetry-focused architectural designs, as exemplified by models such as EGNN[25], SchNet[26], and Geodesic-GNN[27]. While computational scientists, (when first entering into a new field such as drug design) would tend to tinkle with model architectures in order to attain better performance under the existing practices (for instance, a given geometric protocol), it is crucial to recognize that the protocol itself should also be re-assessed if a substantial breakthrough is the goal. The selected protocol sets the performance boundary of a model and significantly dictate the outcome. Therefore, we advocate that a thorough review and re-thinking of existing geometric handling protocols are imperative.

In light of these observations, we first review and summarize six protocols that could be used in 3D molecular generation, highlighting their respective challenges and discussing their usages in other molecular geometry-centric problems, like molecular conformation generation and docking problems. Building on this foundation, we propose a hybrid approach that employs multiple protocols and effectively draws upon the unique strength of each one to achieve an optimal performance in 3D molecular generation, as highlighted in **Figure 2C**. This novel strategy led to the development of the first geometry-reliable and fragment-wise molecular deep generation, FragGen as presented in **Figure 2B**. It achieves state-of-the-art performance in our reported experiments and validating our argument on the need to re-formulate the geometry handling protocol. Furthermore, we grounded our algorithmic development into the real-world drug design campaigns, successfully designing potent type II inhibitors (75.9 nM) targeting the leukocyte receptor tyrosine kinase. This concept-algorithm-application work not only fills a gap in structure-based drug discovery but also enriches the discourse on geometric handling protocols, complementing symmetric neural network design and offering a blueprint for model development for other geometry-related fields.

## Results and Discussions

**Analysis of Geometry Handling Protocols**

| | Initial State | Intermediate | Goal | Challenge | Example | Other Models |
|---|---|---|---|---|---|---|
| Internal Coordinate | | | | | GraphBP | G-SphereNet (MG) GraphVF (S-MG) ... |
| Cartesian Coordinate | | | | | DiffBP | DiffSBDD (S-MG) GeoDiff (CG) ... |
| Relative Vector | | | | | ResGen | Pkt2Mol (S-MG) PocketFlow (S-MG) ... |
| GeomGNN | | | | | FragGen-GNN | KarmaDock (S-CG) EquiBind (S-CG) ... |
| GeomOPT | | | | | FragGen-OPT | Classics |
| Distance Geometry | | | | | FLAG | ConfVAE (CG) SDEGen (CG) ... |
| Combined Strategy | | | | | FragGen | — |

**Figure 3.** This figure presents a comparative illustration of workflows, challenges, objectives, and implementations across different geometry handling protocols. The 'Example' column focuses on applications within the field of 3D molecular generation, while the 'Other Models' column spans a broader range of geometry-centric topics. Key abbreviations include MG: Molecular Generation (without structures), S-MG: Structure-based Molecular Generation, CG: Conformation Generation (without structures), and S-CG: Structure-based Conformation Generation (also known as Docking).

The continuing advancement of structural predictions for various biomolecules, exemplified by AlphaFold, has drawn the AI community's attention onto structure-based drug design, where accurately modeling molecular geometry plays a pivotal role in estimating drug-target interactions. In this context, we meticulously examine six universal geometry handling protocols, as depicted in **Figure 3**, underscoring the unique challenges each of them encounters in the context of pocket-aware 3D molecular design.

The **Internal Coordinate** protocol, which initially determines three atomic orders before predicting bond lengths, angles, and dihedral angles, often leads to distorted molecular conformations. This protocol is adopted by the GraphBP method (**Figure 3**), whose errors have been found to predominantly arise from incorrect determination of the initial topological order, which is inherently difficult to determine within protein pockets. Unlike structure-free models like G-SphereNet[28], where topological orders naturally follow generation trajectories in the ligand-only scenarios, the application of Internal Coordinate protocol in pocket-aware context struggles in the more complex environments, such as the protein pockets. In contrast, the **Cartesian Coordinate** approach, which involves probabilistic learning directly on 3D coordinates, lacks local structural constraints. This often results in the accumulation of errors at each atomic position, leading to implausible geometries, such as non-coplanar rings or benzene rings with unequal bond lengths (**Figure 3**). This challenge is prevalent in diffusion model-based methods like DiffBP and DiffSBDD[29], which generate molecules in one shot. The **Relative Vector** protocol, predicting coordinate vector differences between atoms, appears more robust. Ensuring that the predicted 3D vector satisfies SE(3)-equivariance, this method effectively confines the degrees of freedom to bond lengths, thereby minimizing the impact of prediction errors on overall geometry. Methods like Pocket2Mol and ResGen, which employ this protocol, have achieved more rational generation of conformations. However, they still face challenges, particularly in generating multi-fused ring molecules that, while favoring stronger protein pocket interactions, are complex and difficult to synthesize.

The **GeomGNN approach**, utilized in KarmaDock, leverages equivariant graph neural networks to learn atomic forces, followed by a coarse coordinate update ($x_i = x_{i-1} + F_i$). This protocol benefits from straightforward training and inference, as it avoids complex transitions between different coordinate descriptions. Our implementation in the 3D molecular generation problem, resulting in FragGen-GNN, demonstrates this advantage. However, it also exhibits limitations in achieving precise atom localization. **GeomOPT**, a classical method for determining next atom or fragment coordinates, theoretically avoids local structure implausibility through force-field interactions involving bond angles and dihedrals. Despite its potential, this protocol faces significant limitations, including lengthy optimization times and a tendency for structures to become trapped in local minima, leading to twisted molecular structures, as shown in **Figure 3**. **Distance Geometry**, another recognized approach used by models in conformation generation, such as ConfGF[30] and SDEGen[19], circumvents equivariance demands in neural network design by modeling interatomic distances. This reduces model construction complexity but suffers from an overabundance of degrees of freedom, making it impossible to uniquely determine 3D coordinates from a distance matrix. Consequently, even with a perfectly predicted distance matrix, accurate reconstruction of original Cartesian coordinates remains elusive, often resulting in distorted conformations, as seen with the FLAG method (**Figure 3**).

While ongoing advancements in model architecture design strive for improved performance, they do not directly address the inherent challenges of each geometry protocol summarized above. Recognizing this lack of algorithmic development on an equally important issue that contributes to the overall quality of generated conformations, this work sets out to improve the existing protocol and propose a **Combined Strategy** which integrates insights emerged from our systematic investigation on the pros and cons of each existing protocol.

More specifically, the Combined strategy works as follows. We first utilize the Relative Vector protocol for sub-pocket detection, determining suitable locations for subsequent fragment assembly. Upon predicting the next fragment type, its geometry is decomposed into local conformation, rotation around a point (connected atom), and rotation around an axis (connected bond). Traditional methods and deep learning approaches generally perform well for local fragment geometries. For rotations around a point, we apply hybrid orbital theory constraints[31], such as the consistent bond angles in standard SP3 hybridization (e.g., 109.5° in methane), to guide the molecular assembly with chemical initialization founded on rigorous theoretical insights. Finally, for rotation around an axis, we directly predict dihedral angles using von Mises loss, more details can be found in Method part This decoupling of complex fragment-wise generation geometry has led to an effective solution, with subsequent experiments providing strong validation of our approach.

**Performance of FragGen on the CrossDock Benchmark**

Leveraging our novel geometry handling protocol, we developed FragGen, a structure-based, fragment-wise molecular generation method. Its efficacy was rigorously tested using the widely recognized CrossDock dataset[32], a benchmark in previous atom-wise molecular generation research[12-16]. The evaluation involved calculating the Vina Score with AutoDock Vina[33] to gauge the ligand's binding affinity to its target protein. Additionally, other critical metrics are also included, such as the Quantitative Estimation of Drug-likeness (QED)[34], Synthetic Accessibility (SA)[35], Lipinski's Rule of Five[36], and the octanol-water partition coefficient (LogP), to characterize the properties of the molecules generated. Notably, SA emerged as a crucial metric in contrasting atom-wise and fragment-wise methodologies, with the latter typically yielding higher SA due to the assembly of existing commercial fragments. Our baseline models included four atom-wise molecular generation approaches (GraphBP, DiffBP, Pkt2Mol, and ResGen) and one fragment-wise model FLAG, the only open-source model of its kind. The performance metrics for each model are detailed in **Table 1**.

Table 1. The mean binding energies and drug-like properties for Top1/5 molecules.

|  | Test Set | GraphBP | DiffBP | Pocket2Mol | ResGen | FLAG | FragGen |
|---|---|---|---|---|---|---|---|
|  | Top1 | | | | | | |
| Vina Score (↓) | -7.158 | -9.332 | -9.237 | -9.247 | -9.622 | -8.954 | -9.926 |
| Hit Pocket | - | 87.07% | 9.42% | 92.10% | 93.15% | 87.14% | 96.15% |
| QED (↑) | 0.531 | 0.560 | 0.479 | **0.562** | 0.536 | 0.552 | 0.541 |
| SA (↑) | 0.730 | 0.464 | 0.411 | 0.341 | 0.307 | 0.565 | 0.740 |
| Lipinski (↑) | 4.684 | 4.821 | 4.734 | 4.921 | **4.958** | 4.955 | 4.871 |
| LogP | 0.947 | 1.552 | 0.452 | 0.8249 | 1.891 | 0.746 | 0.154 |
|  | Top5 | | | | | | |
| Vina Score (↓) | -7.158 | -8.515 | -8.723 | -8.924 | -9.343 | -8.188 | -9.654 |
| QED (↑) | 0.531 | **0.563** | 0.492 | 0.571 | 0.546 | 0.522 | 0.573 |
| SA (↑) | 0.730 | 0.478 | 0.433 | 0.346 | 0.316 | 0.582 | 0.717 |
| Lipinski (↑) | 4.684 | 4.776 | 4.788 | 4.931 | 4.953 | 4.975 | 4.859 |
| LogP | 0.947 | 1.430 | 0.457 | 0.758 | 1.646 | 0.451 | 1.273 |

From the results in **Table 1**, FragGen outperforms other methods in Vina Score, ranking as follows: FragGen > ResGen > Pkt2Mol > GraphBP > DiffBP > FLAG. FragGen leads with a Vina Score 2.5 kcal/mol higher than the test set average, translating to over 100-fold increase in binding affinity based on the thermodynamic principles[15]. This significant boost in binding potency is almost enough to elevate a ligand from μM $IC_{50}$ to nM $IC_{50}$. Furthermore, FragGen excels in generating high-quality ligands with superior chemical and geometric structures. As illustrated in **Figure 1**, atom-wise methods like GraphBP and DiffBP often yield distorted molecular geometries, with some GraphBP-generated molecules even straying out of the target pockets. These flawed geometries stem from the limitations of the Internal Coordinate and Cartesian Coordinate protocols, where the latter necessitates predefined topological atomic orders, and the former lacks local structural constraints to guide the generative process. In contrast, ResGen and Pkt2Mol, employing the Relative Vector protocol, achieve more accurate and visually rational molecular geometries. FLAG and FragGen, both fragment-wise approaches, turn out to give outputs that sits on opposite ends of the Vina Score spectrum (FLAG: ~-8.9 vs. FragGen: ~-9.9), a testament to their geometry handling capabilities. FLAG, based on Distance Geometry, often struggles with ill-structured molecules due to the challenges in mapping an extensive number of pairwise distances to Cartesian coordinates. Conversely, FragGen employs a sophisticated geometry handling approach, decomposed into four geometric variables and effectively managed through a blend of chemical knowledge and end-to-end learning. The Vina Score ranking thus basically serves as an indicator of a model's geometry handling proficiency.

Regarding molecular properties, FragGen achieves the highest scores in QED and SA on the Top-5 results, underscoring the chemical viability of its generated molecules. These impressive results stem from two key factors: the inherent nature of the fragment-wise protocol and the advantages of a robust geometry handling approach. The fragment-wise protocol inherently guarantees better synthesizability, as it typically decomposes molecules into a set of existing fragments, also explaining FLAG's relatively high SA score. In contrast, atom-wise methods like Pkt2Mol and ResGen often generate molecules that completely fill the cavity of protein pockets, resulting in lower QED and SA scores. This tendency has contributed to the hesitancy among medicinal chemists to integrate previous molecular generation methods into their workflows. In summary, the advancements of FragGen in terms of Vina Score, QED, and SA indicate that geometric accuracy plays a crucial role in enhancing chemical plausibility, as the geometry of the current molecular state influences the structure of the subsequent fragment. For real-world applications, FragGen also establishes it as a valuable tool in drug discovery, particularly for generating easily synthesizable samples.

**Performance of FragGen on Well-Studied Pharmaceutical Targets**

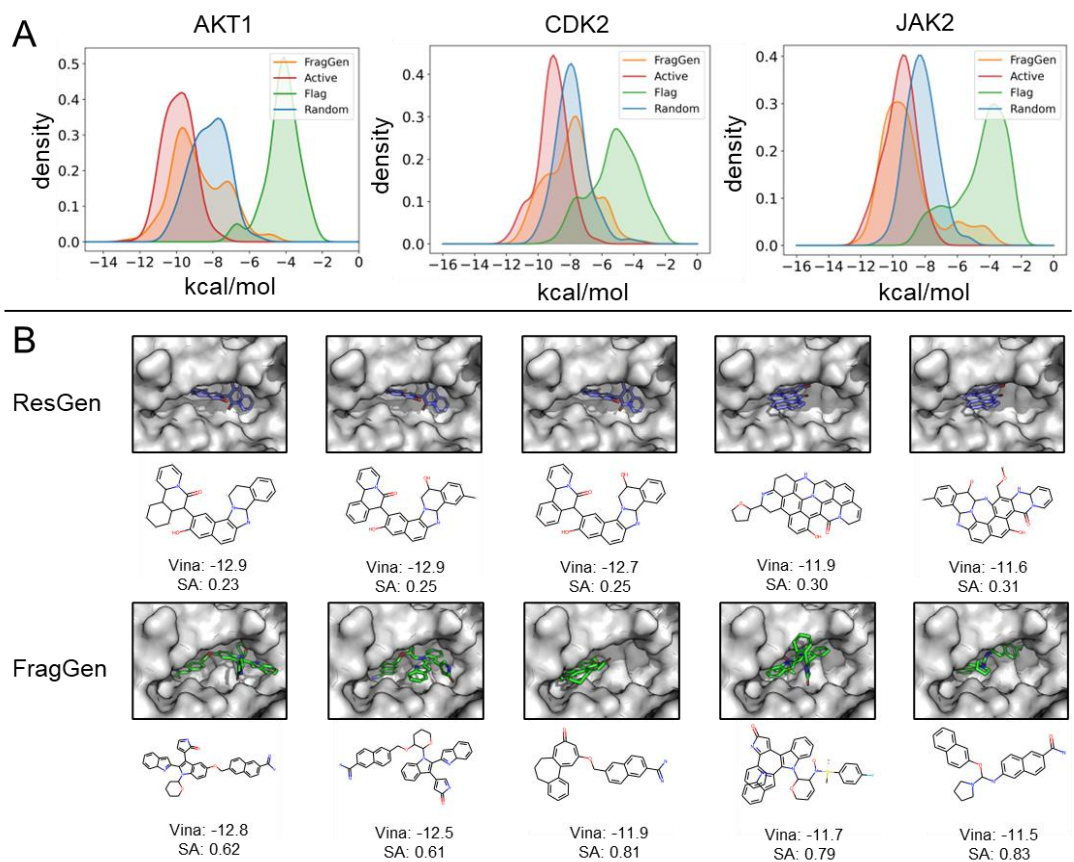

**Figure 4. A).** Distribution of Binding potency (Vina Score) for FragGen and its counterpart across three well-studied targets. B). Comparative visualization of the top 5 molecules in terms of binding potency, highlighting differences between the atom-wise (ResGen) and fragment-wise (FragGen) approaches.

To demonstrate FragGen's applicability in real-world scenarios, we evaluated its performance on several well-studied pharmaceutical targets. These targets, with well-characterized active sites and numerous experimentally discovered inhibitors, provide a suitable testing ground. Unlike the CrossDock benchmark, this experiment included two additional molecule sets: Active (experimentally validated molecules serving as positive controls) and Random (randomly selected chemical moieties from the GEOM-Drug set[37], serving as negative controls). The Vina Score and molecular properties, akin to those used in the CrossDock experiment, are detailed in **Table S1**. **Figure 4A** illustrates the binding potency distribution of FragGen-generated molecules (in orange) in comparison to the fragment-based counterpart, FLAG (in green). Notably, FragGen's distribution aligns more closely with the Active molecules, while FLAG aligns with the Random set. This result again highlights the advantage of a rational geometry protocol in fragment-wise molecular generation, where accurate geometries lead to a better energy match with the binding protein.

From **Table S1**, it is evident that ResGen, a state-of-the-art (SOTA) atom-wise molecular generation method, scores highly in terms of binding potency on targets like AKT1 and CDK2, with FragGen closely following. Despite this, we assert FragGen's superiority, as illustrated in **Figure 4B**. While ResGen's top-generated molecules exhibit strong binding potency, they compromise on synthesizability and drugability. In contrast, FragGen's molecules not only achieve comparable binding potency to the top-Active molecules (with a marginal ~0.4 kcal/mol difference) but also maintain the highest chemical accessibility, making them more favorable for chemists. This is further supported by the SA comparison in **Table S1**, where FragGen outperforms other models.

**Applying FragGen to design Type II inhibitors of LTK with wet-lab validations.**

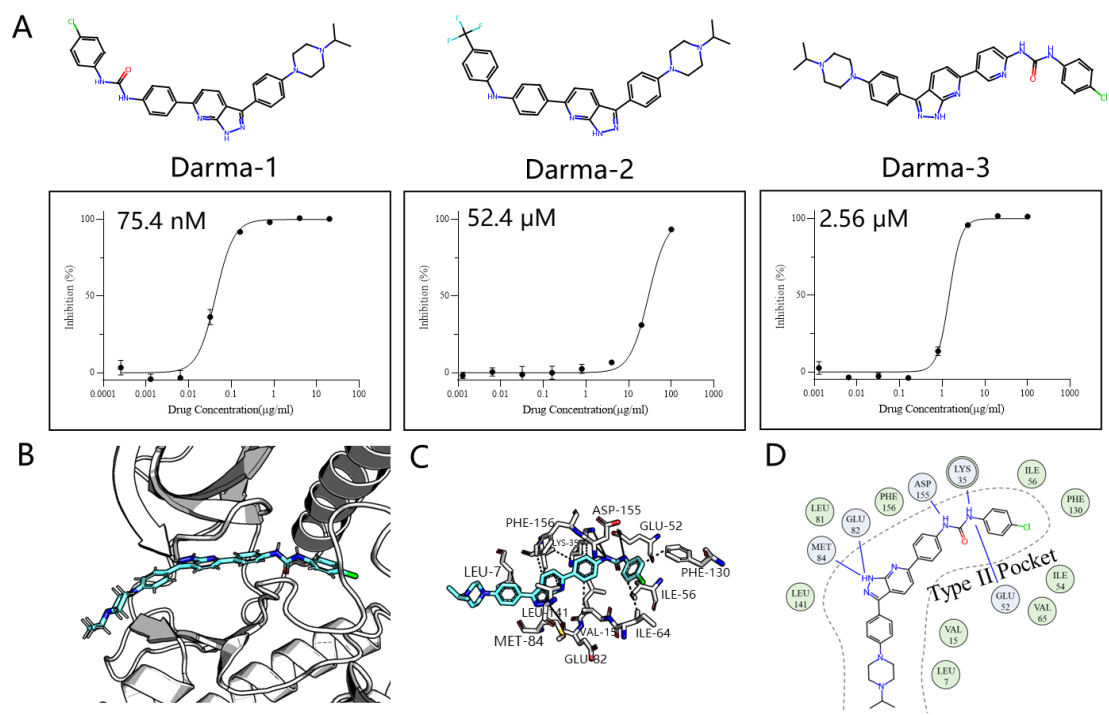

**Figure 5. A)** Structures of the three synthesized compounds designed by FragGen and their inhibitory activity (IC$_{50}$) against Ba/F3-CLIP1-LTK cells. **B)** The binding conformation of Darma-1 in LTK DFG-

out model. **C)** 3D protein-ligand interactions analyzed by PLIP[38]. **D)** 2D visualization of protein-ligand interactions, where the green represents hydrophobic interaction and the blue denotes hydrogen bond interaction.

Kinases, essential enzymes in cellular signaling, play a critical role in various physiological processes, including cell growth, differentiation, and metabolism. As a result, numerous kinase inhibitors have been developed and approved for the treatment of diseases such as cancer, cardiovascular disorders, and inflammation[39]. Traditional kinase inhibitors, known as type I inhibitors, target the ATP-binding sites in the active conformations of kinases, offering therapeutic benefits but facing limitations in selectivity and resistance issues. In contrast, type II inhibitors, like sorafenib, target an additional allosteric site, the DFG-out pocket, potentially enabling more selective and less toxic treatments.[40] Despite the advantages of type II inhibitors, existing computational tools, such as quantitative structure–activity relationship (QSAR) and docking screening[41,42], fall short in designing potent molecules beyond the known chemical space, limiting the scope of discovering novel therapeutic agents. Therefore, the current molecular generation methods are ideal for filling this gap.

We chose the LTK as the validation system, a promising kinase target for treating non-small cell lung cancer according to the recent study[43]. This choice differs fromm previous retrospective studies, not only because it was validated through wet experiments rather than a controversial docking metric, but also because it is a novel target with few inhibitors designed for it. Inspired by the historical drug development of PDGFRβ target, which designs type II inhibitors based on the type I framework[44], we developed an AI-powered structure-based workflow using FragGen. Specifically, we first built the LTK DFG-out homology model based on the anaplastic lymphoma kinase (ALK)[45] protein, owing to their high sequence similarity. Then we docked a previously reported type I inhibitor[46] of ALK into the LTK model, aiming to anchor the molecule at the pocket I region by retaining the head hinge-binding moiety. Starting with the anchored structure, FragGen was utilized to explore the chemical space targeting type II pocket. Within 10 minutes, FragGen proposed 97 chemical candidates. Subsequently, four filtering criteria were applied to narrow down the candidates: 1) number of hydrogen donors < 5; 2) number of hydrogen acceptors < 10; -3) 2 < LogP < 5; 4) and number of rotatable bonds < 10. Out of this group, 10 molecules satisfied these conditions. Among them, three were chosen for further investigation based on synthesis feasibility as recommended by organic chemists (Figure 5A). Details on the synthetic routes and molecular characterization are provided in the Supplementary Information. Bioassays demonstrated high affinities for LTK, with Darma-1 exhibiting notable potency at 75.4 nM. The other two candidates showed affinities of 52.4 μM and 2.56 μM, respectively, highlighting FragGen's ligand design capability within protein pockets. The successful design of potent type II inhibitors may be attributed to FragGen's sophisticated handling of geometries. To illustrate this point, we analyzed the binding mode of the directly generated Darma-1 compound in Figures 5B, 5C, and 5D. It is evident that the generated compound forms comprehensive physical interaction with the type II pocket, like three hydrogen bonds with the ASP-155, LYS-35, and GLU-52 residues. Molecular generation models would lose practical utility if the generated geometries are not as reasonable as those proposed by FragGen no matter how promising the docking metric/ADMET metric they score: improper conformations will disrupt the interaction between proteins and ligands, diminishing the credibility of the generated samples.

**Geometric Plausibility of Generated Molecules**

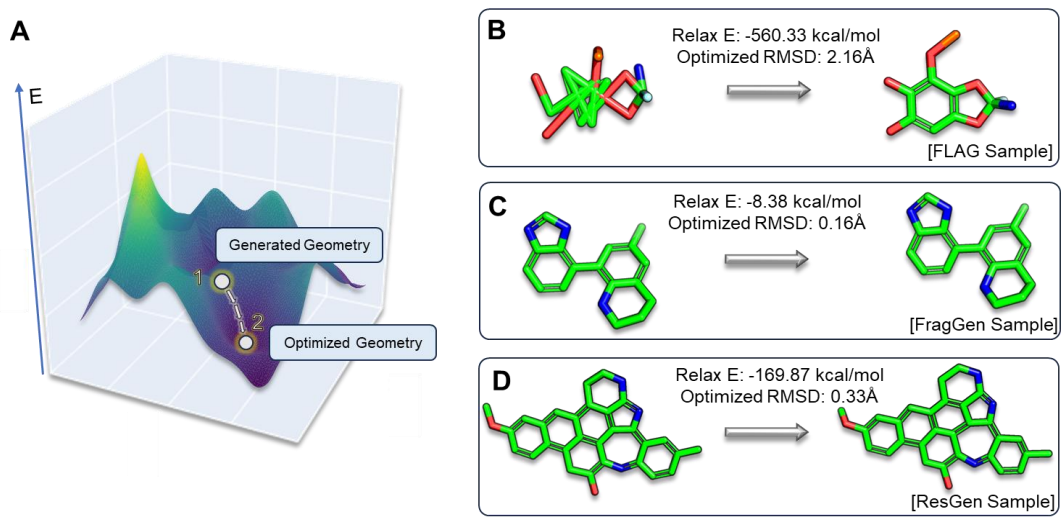

**Figure 6. A).** Visualization depicting the positions of generated and optimized geometries within the energy landscape. **B,C,D).** Three case studies showcasing Relax E and OptRMSD metrics, each illustrating distinct scenarios encountered in FLAG, FragGen, and ResGen.

In the realm of 3D molecular generation, many models rely on resort to geometry optimization to rectify distortions in generated molecules, essentially obscuring the limitations of deep learning methods in co-designing molecules with accurate geometries. Recognizing that previous experiments have only been able to indirectly and qualitatively address these geometric challenges, we introduce two novel metrics to gain a more detailed and quantitative assessment: relaxation energy (Relax E) and optimized root mean square deviation (OptRMSD). Specifically, the generated molecules undergo force field optimization, then the energy released and RMSD between the directly generated and optimized molecules are calculated, as shown in **Figure 6A**.

**Table 2.** The results of OptRMSD and Relax E across different methods.

| Case | GraphBP | DiffBP | Pkt2Mol | ResGen | FLAG | FragGen |
|---|---|---|---|---|---|---|
| OptRMSD | 1.359 ±0.722 | 1.158 ±2.378 | 0.499 ±0.404 | **0.465 ±0.319** | 1.379 ±0.855 | 0.878 ±1.010 |
| Relax E | -83.22 ±288.5 | -100.9 ±235.1- | -46.76 ±40.05 | -54.33 ±45.21 | -387.1 ±481.9 | **-40.26 ±71.45** |

**Table 2** presents the results for Relax E and OptRMSD. Notably, in the realm of OptRMSD, certain models exhibit superior performance. However, it is crucial to acknowledge that OptRMSD inherently exhibits a preference for multi-ring structures. This is due to the fact that larger aromatic systems, with their more rigid frameworks, are less prone to conformational alterations, a phenomenon illustrated in **Figure 6D**. Consequently, the lower OptRMSD scores observed in models like ResGen and Pkt2Mol, which are predisposed to generating multi-ring molecules, align with expectations. In contrast, FragGen distinguishes itself by achieving an OptRMSD score below 1 Å, underscoring its proficiency in creating structurally coherent molecules. When considering Relax E, a metric less biased towards multi-ring structures, a different picture emerges. Multi-ring structures, as shown in **Figures 6C-D**, tend to release more energy following force-field optimization, even when they exhibit similar OptRMSD values to simpler molecules. In this context, FragGen again demonstrates superior performance, effectively aligning with our earlier assessments of its geometric accuracy. Conversely, the fragment-wise method FLAG, along with models like DiffBP and GraphBP that are prone to generating distorted conformations, give less favorable results in this metric.

**Ablation Study of Geometry Handling Protocols in FragGen**

In the 3D molecular generation task, four of the six protocols in **Figure 3**, Internal Coordinate, Cartesian Coordinate, Relative Vector, and Distance Geometry, have been instantiated by works like GraphBP, DiffBP, ResGen, and FLAG, respectively. In addition to these, we have integrated the GeomGNN and GeomOPT protocols into FragGen, creating two more versions of FragGen thereby providing a comprehensive analysis of each protocol within the context of 3D molecular generation. The results of this ablation within FragGen are detailed in **Table S2**.

**Table S2** reveals that molecules generated using the GeomGNN protocol exhibit the highest binding propensity. However, this favorable binding tendency comes at a cost to their synthesizability, which is approximately 24% lower compared to the other protocols. This reduction in synthesizability can be attributed to the compromise in local structural rationality while the model attempts to fill the protein pocket cavity (as depicted in **Figures S1A**) without explicitly considering the overall synthetic feasibility of the molecules. On the other hand, the GeomOPT approach shows a marked improvement in synthesizability, but the molecules generated under this protocol demonstrate a reduced binding tendency. This is primarily due to the geometric conformations becoming trapped in local minima within the protein structure during the generation process, leading to suboptimal molecule-protein interactions, as illustrated in **Figure S1A**. The Combined Strategy, which synergizes the physical constraints and the strengths of both Relative Vector and Internal Coordinates, emerges as a robust approach. It not only facilitates realistic molecule generation but also ensures a potent binding affinity to target proteins. The molecules produced under this strategy not only exhibit a higher binding tendency, outperforming all baseline methods (both atom-level and fragment-level) as shown in **Table 1**, but also demonstrate the highest level of synthesizability among all the protocols. This underscores the effectiveness and rationality of the molecular structures generated through this comprehensive protocol.

## Conclusion

In this study, we aimed to address the frequently encountered issues of implausible chemical and geometric structures generated by many 3D molecular generative models. This journey began with a meticulous identification and analysis of six geometry handling protocols, each with its unique strengths and shortcomings. After acquiring the insights on the problems associated with existing approaches, we proposed developed FragGen, a hybrid strategy tailored for structure-based fragment-wise molecular generation. Experiments across the recognized benchmark and pharmaceutically relevant targets demonstrate that FragGen-generated molecules exhibit the highest binding potency (as estimated with docking scores) and synthesizability, meeting the practical demands of real-world drug discovery efforts. Our detailed geometric analysis and ablation study demonstrate that FragGen effectively coordinates the intricate interplay between molecular geometry and protein pocket structure, highlighting the crucial role of our proposed hybrid strategy in combining various geometry handling techniques to achieve FragGen's remarkable success. Finally, we successfully employed FragGen to design potent LTK type II inhibitors, showcasing its practical utility and completing the final step in the concept-algorithm-application chain. In summary, by integrating insights from different geometry handling protocols and tailoring them to the specific needs of fragment-wise molecular generation, FragGen has emerged as a robust tool for structure-based drug design. Its ability to generate molecules with high binding affinity, coupled with practical synthesizability and geometric fidelity, not only adds an applicable tool to chemists' toolboxes, but also serves as a well-principled method that can recommend suitable geometry handling for a broader range of geometry-related problems in deep learning.

## Methods

**Protein-Ligand Interaction Learning Module**

To fully perceive the protein-ligand interaction, we first construct the protein-ligand graph and then apply the geometric message passing framework to them. This framework is described in the following formula:

$$\left(n'_{p_i}, \vec{n}'_{p_i}\right) = \text{Emb}(n_{p_i}, \vec{n}_{p_i}),$$

$$\left(n'_{l_i}, \vec{n}'_{l_i}\right) = \text{Emb}(n_{l_i}, \vec{n}_{p_i}),$$

$$(h_i, \vec{h_i}) = \text{GeomEncoder}(n_{l_i}, n_{p_i}, \vec{n}_{l_i}, \vec{n}_{p_i}, e_{ij}, \vec{e}_{ij}).$$

where $n_p$ and $n_l$ denote the node features of proteins and ligands; $\vec{\ }$ signifies the vector features; $e_{ij}$ is the edge features between nodes $i$ and $j$; $h_i$ refers to the hidden features of the protein-ligand graph. Emb is the embedding layer, which maps the raw features of protein and ligand to the corresponding spaces with the same dimension. GeomEncoder is composed of several interaction layers based on geometric equivariant networks. The detailed architectures of Emb and GeomEncoder can be found in Part 1, SI.

**Frontier Prediction**

To autoregressively generate the subsequent fragment, it is crucial to predict the frontier atom within the existing ligands. Notably, at the initial stage, there are no ligand atoms present, so the frontier is chosen from among the protein atoms. The probability of selecting the frontier from either ligand atoms or protein atoms can be simplified and represented as follows:

$$\left(n_{f_i}, \vec{n}_{f_i}\right) = f\left(\text{SL}_{f1}(h_i), \text{VL}_{f1}(\vec{h_i})\right),$$

$$p_{f_i} = \sigma\left(\text{SL}_{f3}\left(\|\vec{n}_{f_i}\|_2 + f\left(\text{SL}_{f2}\left(n_{f_i}\right)\right)\right)\right).$$

where $p_{f_i}$ is the focal probability of node $i$; $\sigma$ is the sigmoid function; SL and VL denote scalar layers and vector layers[47], respectively. $n_{f_i}, \vec{n}_{f_i}$ are intermediate scalar and vector features.

**Cavity Detection**

Once the frontier has been established, the next step is to predict the cavity where the subsequent fragment can be optimally positioned. This prediction of the next cavity is accomplished using a mixture density network, which is implemented as follows:

$$\left(r_i, \vec{r_i}\right) = \text{GVP}_r\left(\text{SL}_{x1}(h_i), \text{VL}_{x1}\left(\vec{h_i}\right)\right),$$

$$\left(w_i, \vec{w_i}\right) = \text{GVP}_w\left(\text{SL}_{x2}(h_i), \text{VL}_{x2}\left(\vec{h_i}\right)\right),$$

$$\left(\Sigma_i, \vec{\Sigma_i}\right) = \text{GVP}_\Sigma\left(\text{SL}_{x3}(h_i), \text{VL}_{x3}\left(\vec{h_i}\right)\right),$$

$$\vec{x_i} = \vec{x}_{ai} + \sum_{k=1}^{K} w_i^k \vec{r}_i^k.$$

where $\vec{r_i}$ is the predicted relative vector, $w_i$ and $\Sigma_i$ are the factor and variance of the $i$-th component of the mixture Gaussian density, respectively, $\vec{x}_{ai}$ is the coordinate of the focal atom, and $\vec{x}_i$ is the detected cavity coordinate. GVP is the geometric vector perceptron[48], which can be found in Part 1, SI.

**Fragment Query**

Once the next cavity is identified, we can begin to search for suitable fragments that can be placed within it. It is important that the placement adheres to the principles of geometry and energy matching, which requires a thorough understanding of the local cavity environment. To achieve this, we gather detailed information about the cavity. This data is then integrated with the frontier features to facilitate an informed query for the appropriate fragment placement:

$$y_{m_{ij}}, \vec{y}_{m_{ij}} = \text{GeomMessage}\left(h_i, \vec{h}_i, e_{ij}, \vec{e}_{ij}\right),$$

$$y_{h_i}, \vec{y}_{h_i} = \sum_{k=1}^{j} \left(y_{m_{ik}}, \vec{y}_{m_{ik}}\right),$$

$$p_{y_i} = \sigma\left(SL_{t2}\left(\parallel \vec{y}_{h_i} \parallel_2 + f\left(SL_{t1}(y_{h_i})\right)\right)\right).$$

where $y_{m_{ij}}, \vec{y}_{m_{ij}}$ are the message between $i$, cavity node, and $j$, the K nearest neighborhoods of node $i$. $y_{h_i}, \vec{y}_{h_i}$ are clustered type hidden features on the cavity node $i$, and $p_{y_i}$ is the probability of the next fragment type. GeomMessage is the message block that makes cavity node $i$ blended with its pocket environment.

**Attachment Selection**

The key difference between atom-wise and fragment-wise generation lies in the uncertainty associated with selecting the appropriate atom within a predicted fragment for connection and determining its subsequent geometry. Methods like FLAG addresses this challenge by pre-storing fragments with annotated connection points. While effective, this approach significantly increases the size of the fragment database and lacks elegance. In contrast, FragGen directly addresses this challenge using a Graph Attention Network (GAT)[49], a two-dimensional approach, to extract chemical information from the upcoming fragment. Additionally, a geometric network is applied to the frontier node to gather geometric information, such as the influence of existing molecular states and their interaction with protein pockets on the selection of the attachment point. This innovative approach is operationalized as follows:

$$h_{a_i}, \vec{h}_{a_i} = \text{GVP}_{\text{atta}}(h_i, \vec{h}_i),$$

$$h'_{f_j} = \text{GAT}(h_{f_j}, e_{f_j}),$$

$$y_{cr}^{emb}, y_{nx}^{emb} = \text{Embed}(y_{cr}, y_{nx}),$$

$$h'_{a_j} = (h'_{f_j} \| y_{cr}^{emb} \| y_{nx}^{emb} \| h_{a_i}),$$

$$p_{a_j} = \sigma\left(\text{MLP}\left(h'_{a_j}\right)\right).$$

where $h_{a_i}, \vec{h}_{a_i}$ are the hidden features of i-th node's connected atom, i.e., focal atom; and $h'_{f_j}$ are the hidden feature of next fragment's atom $j$; $h_{f_j}, e_{f_j}$ are atom and edge features within the next fragment, respectively; $y_{cr}, y_{nx}$ are the current and next fragment types, respectively, and $y_{cr}^{emb}, y_{nx}^{emb}$ are their corresponding embeddings; $h'_{a_j}$ is the concatenated feature of j-th atom in next fragment, and $\|$ is the concatenate operation; and $p_{a_j}$ is the probability of the attachment of j-th node in the next fragment.

**Bond Linking**

After identifying the next attachment atom, the subsequent variable to predict is the covalent bond. While many molecular generation methods, such as DiffSBDD, determine bonding relationships using empirical rules, FragGen takes a direct prediction approach that is both valence- and geometry-aware. The reason for incorporating geometric considerations is that the local pocket environment may favor certain types of interactions, such as the formation of π-π stacking interactions. At the same time, valence constraints guide bond prediction, ensuring that the cumulative valence from forming bonds does not exceed the valence capacity determined by the valence states of the two connected atoms. These principles are operationalized as follows:

$$h_{b_i}, \vec{h}_{b_i} = \text{GVP}_{\text{bond}}(h_i, \vec{h}_i),$$

$$h_{d_{ij}}, h_{nx} = \text{MLP}(d_{ij}, n_{nx}),$$

$$y_{cr}^{emb}, y_{nx}^{emb} = \text{Embed}(y_{cr}, y_{nx}),$$

$$h_{\text{valen}} = \text{MLP}(valen_{cr} \| valen_{nx}),$$

$$p_{b_{ij}} = \sigma\left(\text{MLP}\left(h_{b_i} \| h_{d_{ij}} \| y_{cr}^{emb} \| y_{nx}^{emb} \| h_{\text{valen}}\right)\right).$$

where $h_{b_i}, \vec{h}_{b_i}$ are the features of bonded atom, i.e., focal atom; $d_{ij}$ is the distance between focal node i and cavity node j; $n_{nx}$ is the bonded atom of next fragment; $valen_{cr}$ and $valen_{nx}$ are valence of current and next bonded atoms, respectively; $h_{valen}$ is the concatenated feature of valance information; and the $p_{b_{ij}}$ is the probability of bond type between the current and next bonded atoms i and j.

**Chemical Initialization**

As mentioned earlier, the geometry of the next fragment can be divided into four components. For the local geometries and rotation around the point, the former can be effectively achieved by the DL approach or a classical approach, as exemplified in the SDEGen[19], and the latter benefits from an end-to-end approach. In our novel approach, we integrate knowledge from hybrid orbital theory, which has been instrumental in elucidating molecular conformations, into our prediction process. To illustrate this, consider a methane fragment; it naturally adopts a tetrahedral structure, thereby fixing the rotation around the point. When predicting the conformation of such a fragment, we first identify its connection to the existing molecule via a predicted bond. This involves defining a vector from the focal atom to the next attachment point (the to-be-aligned vector) and another from the focal atom to a designated pocket node (the target vector). We then compute a rotation matrix that aligns these vectors. This matrix is applied to rotate the fragment's conformation, initially set in a vacuum, to establish the initial geometry of the next fragment. The computation of this matrix proceeds as follows:

$$a_{\text{norm}} = \frac{a}{||a||}$$

$$b_{\text{norm}} = \frac{b}{||b||}$$

$$v = a_{\text{norm}} \times b_{\text{norm}}$$

$$c = a_{\text{norm}} \cdot b_{\text{norm}}$$

$$[v]_\times = \begin{bmatrix} 0 & -v_z & v_y \\ v_z & 0 & -v_x \\ -v_y & v_x & 0 \end{bmatrix}$$

$$R_{ab} = I + [v]_\times + [v]_\times^2 \frac{1}{1+c}$$

$$r_f' = R_{ab} r_f$$

where $a$ and $b$ are to-be-aligned and the target vectors, respectively, $R_{ab}$ is the rotation matrix from vector $a$ to $b$, $r_f$ is the fragment conformation generated in vacuum, and $r_f'$ is the initialized fragment conformation.

**Dihedral Handling**

For the next geometric variable, rotation around an axis, we employ a direct prediction method. This approach leverages both the geometric information of the connected atoms and the global characteristics of the ligands. The primary objective is to minimize the overall energy while simultaneously avoiding spatial clashes. The process of handling dihedral angles is executed as follows:

$$\left(h_i^{tor}, \overrightarrow{h_i^{tor}}\right) = \text{GeomEncoder}(n_l, n_p, \vec{n}_l, \vec{n}_p, e_{ll,pp,pl}, \vec{e}_{ll,pp,pl}),$$

$$h_{mol} = \sum_{i=1}^{N} h_i^{tor},$$

$$\theta = \text{MLP}(h_a \| h_b \| h_{mol}),$$

$$R(\mathbf{u}, \theta) = \begin{bmatrix} \cos(\theta) + u_x^2(1-\cos(\theta)) & u_x u_y(1-\cos(\theta)) - u_z \sin(\theta) & u_x u_z(1-\cos(\theta)) + u_y \sin(\theta) \\ u_y u_x(1-\cos(\theta)) + u_z \sin(\theta) & \cos(\theta) + u_y^2(1-\cos(\theta)) & u_y u_z(1-\cos(\theta)) - u_x \sin(\theta) \\ u_z u_x(1-\cos(\theta)) - u_y \sin(\theta) & u_z u_y(1-\cos(\theta)) + u_x \sin(\theta) & \cos(\theta) + u_z^2(1-\cos(\theta)) \end{bmatrix},$$

$$r_f'' = R(\mathbf{u}, \theta) r_f'.$$

where $n_l, n_p, \vec{n}_l, \vec{n}_p, e_{ll,pp,pl}, \vec{e}_{ll,pp,pl}$ are the node and edge features of ligand and protein, $ll, pp, pl$ denotes edge within ligands, within proteins, and between them, respectively; $h_{mol}$ is the summation of ligand features; $h_a$ and $h_b$ are the features of current and next bonded atom, respectively, i.e., focal atom and the next attachment atom; $\theta$ is the predicted dihedral angle; $R(\mathbf{u}, \theta)$ is the rotation around the predicted bond vector $(r_a - r_b)$; $r_f'$ is the initialized fragment conformation; and $r_f''$ is the final predicted fragment conformation.

**Loss function**

The total loss function is:

$$\mathcal{L} = -\frac{1}{n}\left(\sum_{i=1}^{n} f_i \cdot \log p_{f_i} + (1-f_i) \cdot \log(1-p_f)\right)$$

$$-\frac{1}{m}\left(\sum_{i=1}^{m} a_j \cdot \log p_{a_j} + (1-a_j) \cdot \log\left(1-p_{a_j}\right)\right)$$

$$-\log \sum_{k=1}^{K} w_i^{(k)} \mathcal{N}\left(x_i^{(k)} + r_{a_i}, \Sigma_i^{(k)}\right)$$

$$-\sum_{i=1}^{n} y_i \log p_{y_i} - \sum_{j=1}^{n} b_{ij} \log p_{b_{ij}}$$
$$- \log\left(\frac{e^{\kappa \cos(\theta - \mu)}}{2\pi I_0(\kappa)}\right).$$

where $f_i$ and $p_{f_i}$ are the frontier atom label and prediction, respectively, and $n$ is the total number of the existing ligand/protein atoms; $a_j$ and $p_{a_j}$ are the attachment atom label and preiction, respectively, and $m$ is the number of the next fragment atoms; $x_i^{(k)}$, $w_i^{(k)}$, $\Sigma_i^{(k)}$ are the k-th component of the relative vector, coefficient, and variance in the cavity detection module, respectively, and $K$ is the number of components; $y_i$ and $p_{y_i}$ are predicted fragment label and prediction, respectively; $b_{ij}$ and $p_{b_{ij}}$ are predicted bond label and prediction, respectivly. The final term is the von-mises loss, aiming to evaluate how close are two angles. In this loss, $\mu$ and $\theta$ are dihedral angle label and prediction, respectively, $\kappa$ is the concentration parameter, a higher value means a more peaked distribution, and the $I_0$ is the modified Bessel function of order 0.

**Cell Culture**

Ba/F3 cells are cultured in RPMI M Medium 1640 (U21-279b, YOBIBIO) with 10% FBS (F8318, Sigma-Aldrich) and 10 ng/ml IL-3(90143ES10, Yeasen). 293T cells are cultured in DMEM (U21-265B, YOBIBIO) with 10% FBS. All growth media are supplemented with 1% Penicillin-Streptomycin-Glutamine (10378016, Gibco). Cell cultures are maintained in culture flasks in 5% $CO_2$ atmosphere at 37 °C.

**Transformation of Ba/F3-CLIP1-LTK Cell Line**

CLIP1-LTK fusion genes are generated based on cDNAs of human-derived CLIP1 and LTK genes using pLV vector. 293T cells are transfected with pLV-CLIP1-LTK to produce retrovirus particles. The viral supernatants are collected and concentrated following the instructions of Lenti-X Concentrator (631231, Takara). Ba/F3 cells are subsequently transfected with the virus and selected with 2 μg/ml puromycin to obtain Ba/F3-CLIP1-LTK cell line.

**Ba/F3-CLIP1-LTK Activity Assay**

$1 \times 10^4$ Ba/F3-CLIP1-LTK cells are seeded in 96-well plates with RPMI-1640 and treated with gradient concentrations of interest compounds for 48 h. Afterward, 10 μL of 5 mg/mL MTT solution is added into each well and the cells are further incubated for another 4 h. Then, 100 μL of triplex 10% SDS-0.1% HCl-PBS solution is added to dissolve the formazan deposited on the bottom of the plates, and the plates are then further retained in an incubator overnight. The absorbance at 570 nm is measured with the reference wavelength at 650 nm using a Synergy H1 microplate reader (BioTek).

## Data and Code Availability

The data and source code of this study is freely available at GitHub (https://github.com/HaotianZhangAI4Science/FragGen) to allow replication of the results.

## Acknowledgments

This work was supported by This work was financially supported by National Key Research and Development Program of China (2022YFF1203003), National Natural Science Foundation of China (22220102001), and Natural Science Foundation of Zhejiang Province (LD22H300001).

## Author contributions

O.Z. and Y.H. contributed to the main idea and code; S.C. and P.C. contributed to the bioassays; M.Y. and S.C. contributed to the chemical synthesis; X.Z. and H.T. contributed to the ablation study; Y.Z. and M.W. contributed to the data presentation and data collection; Z.W., H.F., Z.Z., and H.C contributed to the baseline models application; Y.K. and C.-Y.H. contributed to the manuscript envision and experimental design. T.H. contributed to the essential financial support, the conceptualization, and was responsible for the overall quality.

## Competing Interests

The authors declare that there is no conflict of interest.

# Supporting Information

**Part S1**. The detailed architectures of several models. **Part S2**. Additional results of retrospective studies on three well-studied targets. **Part S3.** Ablation study of geometry handling protocols in FragGen. **Part S4.** Synthesis routes and molecular characterization of validated compounds. **Figure S1**. Fragment decomposition of crystal ligand and FragGen's top generated molecules. **Figure S2**. Illustration of ablation studies. **Table S1.** The Top5 molecules mean binding energies and drug-like properties across three well-studied targets; **Table S2.** The ablation results of three geometry handling protocols in FragGen.